\documentclass[aps,twocolumn,showpacs,floats,prl]{revtex4}

\usepackage{graphicx}
\usepackage{dcolumn}
\usepackage{bm}
\usepackage{color}
\usepackage{subfigure}

\usepackage[normalem]{ulem}

\newcommand{\new}[1]{{\textbf{\color{blue}#1}}}

\begin{document}

\author{ A. Macia, J. Boronat \new{ ,} and F. Mazzanti }

\affiliation{
 Departament de F\'{i}sica i Enginyeria Nuclear,
  Universitat Polit\`{e}cnica de Catalunya, Campus Nord B4-B5,
  E-08034, Barcelona, Spain
}

\title{Phase diagram of dipolar bosons in 2D with tilted polarization}

\begin{abstract} 
We analyze the ground state of a system of dipolar bosons moving in
the $XY$ plane and such that their dipolar moments are all aligned in
a fixed direction in space. We focus on the general case where the
polarization field forms a generic angle $\alpha$ with respect to the
$Z$ axis. We use the Path Integral Ground State method to analyze the
static properties of the system as both $\alpha$ and the density $n$
vary over a wide range were the system is stable.  We use the maximum
of the static structure function as an order parameter to characterize
the different phases and the transition lines among them.  We find
that aside of a superfluid gas and a solid phase, the system reaches a
stripe phase at large tilting angles that is entirely induced by the
anisotropic character of the interaction. We also show that the
quantum phase transition from the gas to the stripe phase is of second
order, and report approximate values for the critical exponents.

\end{abstract}

\maketitle 

In recent years, dipolar Bose gases have received much attention.  The
study of quantum degenerate gases of dipolar species has become one of
the most active areas of experimental and theoretical research in the
field of ultracold atoms~\cite{Lahaye_09, Baranov_00,
  Baranov_12}. From the theoretical point of view, the anisotropic and
long range character of the interaction make dipolar systems unique,
exhibiting features like $p$-wave superfluidity in two-dimensional
(2D) Fermi gases~\cite{Bruun08} or roton
instability~\cite{Santos_03,Odell_03, Macia_12}.  These two properties
of the dipole-dipole potential enrich the phase diagram when compared
with other systems with more common isotropic interactions of the Van
der Waals type.  In this way, the realization of systems featuring
strong dipolar forces opens prospects for investigating new and highly
interesting many-body effects, not present in other systems.

Up to now, most of the theoretical work on dipolar systems in two
dimensions have focused on the most simple case where the dipolar
moments are all aligned in the normal direction to the plane where
they move. Less attention has been paid to the more general situation
where dipoles are polarized along an arbitrary direction, including
the analysis of scattering properties~\cite{Ticknor_11} or the
superfluid and collapse instabilities of a quasi-two-dimensional gas
of dipolar fermions aligned by an  external
field~\cite{Bruun_08}.
One remarkable feature induced by the anisotropy of the interaction is
the emergence of a stripe phase, which has been predicted to
appear both in Bose~\cite{Macia_12} and Fermi~\cite{Yamaguchi_10,
  Sum_10, Parish_12}. Some of these calculations were done in the
mean field approximation, predicting the appearance of stripes even in
the isotropic case where all dipoles are polarized perpendicularly to
the plane of movement, although recent Monte Carlo calculations 
arrived to different conclusion~\cite{Matveeva_12}. In this letter 
we address this topic and characterize the existence of the stripe 
phase as a function of the density and polarization angle for the
Bose case, and determine the corresponding solid and gas transition lines.

In previous works we discussed the low-density
properties~\cite{Macia_11} and elementary excitation
spectrum~\cite{Macia_12} of the fully anisotropic 2D dipolar
interaction. In this Letter we extend the analysis and investigate the
phase diagram of a 2D system of bosonic dipoles tilted by an angle
$\alpha$ with respect to the normal direction to the plane where they
move. The tilting angle is assumed to be produced by an external
polarization field that makes all the dipoles tightly point along a
fixed direction in space. We omit any kind of second order
interactions between the dipoles and the polarization field, so the
net effect of the latter is to simply orient them along a fixed
direction in space.

In contrast to the particular $\alpha=0$ case, the dipole-dipole
interaction is in general fully anisotropic This property brings
additional degrees of freedom which make possible a stable stripe
phase. Using first-principles quantum Monte Carlo we have established
in this work the phase diagram of bosonic tilted dipoles in 2D and
characterized the phase transitions between the gas, crystal and
stripe phases.

The model Hamiltonian describing
a system of $N$ polarized and interacting dipoles is written as
\begin{equation}
H = -{\hbar^2 \over 2m}\sum_{j=1}^N \nabla_j^2 
+ {C_{dd} \over 4\pi} \sum_{i<j}^N
\left[ {1 - 3\lambda^2 \cos^2\theta_{ij} \over r_{ij}^3} \right] \ ,
\label{Hamiltonian}
\end{equation}
where $C_{dd}$ is proportional to the square of the (magnetic $\mu$ 
or electric $d$) dipole moment, and
$\lambda=\sin\alpha$. Polar coordinates $(r_{ij},\theta_{ij})$ 
describe the separation between particles 
$i$th and $j$th, respectively. 
In the following we use dimensionless units obtained from the 
characteristic dipolar length $r_0=m C_{dd}/(4\pi\hbar^2)$ and
energy $\varepsilon_0 = \hbar^2/(m r_0^2)$.

We perform stochastic Path Integral Ground State (PIGS)~\cite{Sarsa_00}
calculations in order to build the $T=0$ phase diagram of the system
as a function of the dipolar density $n r_0^2$ and polarization angle
$\alpha$.  We simulate a finite number of particles $N$ in a box of
area $A=N/(nr_0^2)$ with periodic boundary conditions.  One relevant feature,
already present in the two-body problem, is the fact that, in the
absence of additional two-body forces, the system can only be stable
when the dipolar interaction is strictly non-negative.  In this way,
there is a critical tilting angle $\alpha_c \simeq 0.61$ above which
the system collapses because the interaction produces regions where
it becomes attractive. 

The efficiency of the PIGS method is largely
enhanced when a suitable variational wave function is used at the end
points of the chains representing the interacting particles. 
Anyway, it is important to remark that estimations of any observable in PIGS
are unbiased with respect to that trial wave function and that, even without it, 
the results remain unchanged~\cite{Rota_10}. We have
checked that, despite the fact that the interaction is anisotropic,
the trial wave function does not need to explicitly incorporate
that feature, and we have chosen a standard Jastrow product
\begin{equation}
\Psi_T({\bf r}_1, {\bf r}_2, \ldots, {\bf r}_N) = \prod_{i<j} f(r_{ij}) \ ,
\label{Jastrow}
\end{equation}
with the two-body correlation factor given by
\begin{equation}
f(r) = \left\{
\begin{array}{lc}
K_0(2/\sqrt{r}) & \mbox{if\,\,}r\leq R_M \\
B\exp\left[-\left( {C\over r} + {C\over L-r} \right) \right] & \mbox{if\,\,}r>R_M
\end{array}
\ .
\right.
\label{two_body}
\end{equation}
where $B, C$ and $R_M$ are constants to be fixed at each density and
tilting angle, and $L$ is the side of the simulation box. By imposing
$f(L/2)$ to be 1, and $f(r)$ and $f'(r)$ to be continuous at $r=R_M$,
only $R_M$ remains unknown, and we determine its value through a
variational optimization. The two-body correlation factor $f(r)$ 
built in this way reproduces the exact
behavior of the zero-energy solution of the $\alpha=0$ two-body
problem at short distances, matched with the box-symmetryzed form of a
phononic wave function in two dimensions~\cite{Reatto_67}.

We know that for $\alpha=0$ the system remains in gas phase up to a
freezing density $n r_0^2 \sim 290$, where the system undergoes a
first order phase transition to a triangular solid~\cite{Astra_07,
  Pupillo_07}. A similar behavior happens when the polarization angle
increases, although the transition density changes with $\alpha$.  In
our simulation we still use Eqs.~(\ref{Jastrow}) and~(\ref{two_body})
at the end points of the PIGS chains in the crystal phase, but in this
case the starting configuration correspond to the sites of the
triangular lattice that optimally describe the system
This corresponds to an equilateral triangular lattice at $\alpha=0$,
but that changes when $\alpha$ increases, squeezing the fundamental
triangle of the lattice in such a way that the distance between
particles in the direction parallel to the projection of the dipolar
moment on the plane is reduced.
Our results for the deformation angles at each angle $\alpha$ are
compatible with classical Monte Carlo simulations where the
potential energy of the system is minimized.

\begin{figure}
\begin{center}
\includegraphics[width=1.\linewidth]{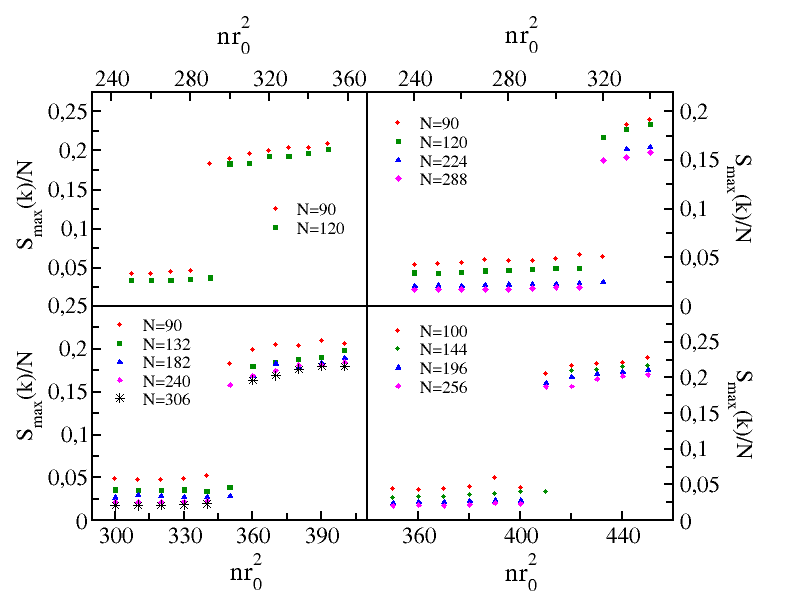}
\caption{(color online). 
Evolution of the order parameter $S_{max}({\bf k})/N$ with the density
for $\alpha=0.1$ (upper left), $\alpha=0.2$ (upper right), $\alpha=0.3$
(lower left) and $\alpha=0.4$ (lower right) for an increasing 
number of particles $N$. 
}
\label{fig_Sk_gas_Crystal}
\end{center}
\end{figure}

In order to characterize the gas-solid transition we use the maximum
strength of the static structure factor divided by the number of
particles in the simulation as an order parameter, $\eta =
S_{max}({\bf k})/N$~\cite{max_Sk}.  Figure~\ref{fig_Sk_gas_Crystal}
shows $\eta$ as a function of the density for the polarization angles
$\alpha=0.1, 0.2, 0.3, 0.4$ and different number of particles in the
simulation. One can clearly see from the figure that up to a certain
density the order parameter is zero, corresponding to a phase where
the main peak in $S({\bf k})$ does not increase significantly with the
number of particles. At higher densities, though, $\eta$ approaches a
constant non-zero value, revealing the existence of a main Bragg peak.
The discontinuity point indicates the transition density at which
crystallization takes place. Table~\ref{Table_gas_Crystal} shows the
transition densities for several tilting angles $\alpha$, while
$\gamma$ stands for the deformation angle, defined in terms of the
primitive vectors of the Bravais lattice
\begin{equation}
{\bf a}_1 = a\hat\imath
\,\,\, , \,\,\,
{\bf a}_2 = {a\over 2}\left( \hat\imath + \hat\jmath \tan\gamma \right)
\label{Bravais}
\end{equation} 
with $a$ fixed by the density. One sees from the table that the
transition density increases with the polarization angle, due to the
fact that, overall, the strength of the interaction decreases when
$\alpha$ increases.  Being the gas-crystal transition of first
order, there are two densities (freezing and melting) defining the coexistence region. In
the current case these two densities must be quite close to each
other as we have not been able to resolve them from our
simulations, as happened also in the isotropic case~\cite{Astra_07}. 
Finally we have checked that the gas-crystal transition
line is well characterized by a parabolic curve of the 
form $n_c r_0^2 = a + b \sin^2\alpha$, with $\alpha=281.75\pm 2.75$
and $b=836.41 \pm 34.38$.

\begin{table}[t]
\begin{tabular}{|c|ccccc|} \hline
$\alpha$ & 0.0 & 0.1 & 0.2 & 0.3 & 0.4 \\  
$\gamma$  & 60$^\circ$ & 60$^\circ$ & 60$^\circ$ & 62$^\circ$ & 64$^\circ$ \\ 
$n_c r_0^2$ & 280(20) & 290(20) & 320(20) & 350(20) & 410(20) \\  \hline
\end{tabular}
\caption{Tilting angle $\alpha$, deformation angle $\gamma$ and transition
densities $n_c r_0^2$ of the gas-crystal transition.}
\label{Table_gas_Crystal}
\end{table}

\begin{figure}
\begin{center}
\includegraphics[width=0.49\textwidth]{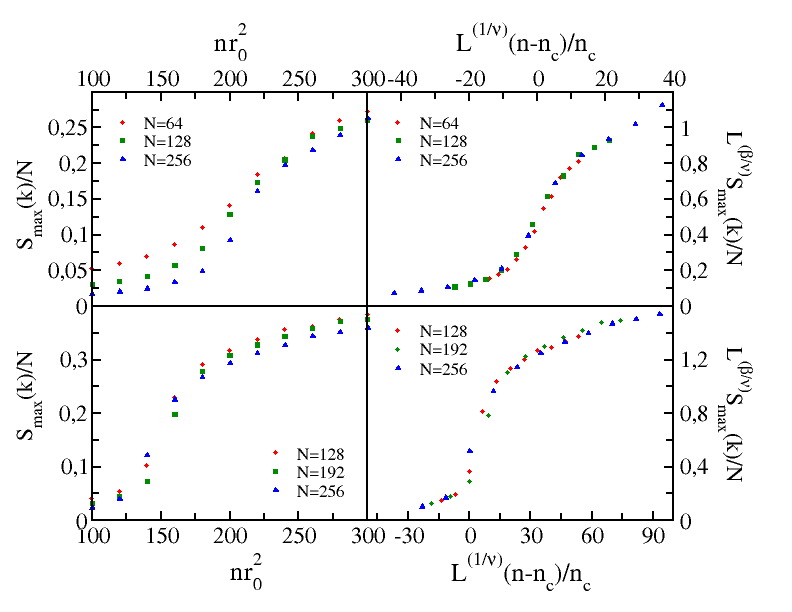}
\caption{(color online). 
Order parameter $\eta$ for $\alpha=0.54$ and $0.58$ as a function of the
dipolar density(upper and lower left panels, respectively).
}
\label{fig_stripes_054}
\end{center}
\end{figure}

By increasing further the tilting angle a new ordered phase appears.
This new phase, characterized by the arrangement of particles in
stripes along the direction where the interaction presents weakest
strength, has been reported to exist in Bose~\cite{Macia_12} and
Fermi~\cite{Yamaguchi_10, Sum_10, Parish_12} systems of 2D dipoles.
The stripe phase is characterized by the emergence of Bragg peaks in
$S({\bf k})$ due to the spatial ordering in one direction compared
with the gas phase.  For this reason we also use here the $\eta$
parameter defined above in order to characterize now the transition
from gas to stripes.  The upper and lower left panels of
Fig.~\ref{fig_stripes_054} show the evolution of $\eta$ with the
density for $\alpha=0.54$ and $0.58$, respectively, and for different
number of particles in the simulation.  The different behavior when
compared with the gas-crystal transition is evident and shows that the
transition is in this case continuous.  As we are always simulating a
finite system in a box with periodic boundary conditions, we use
finite size scaling near the transition point in order to find the
critical exponents of this second-order phase transition. We thus
employ the following form of a length-scaled order parameter
\begin{equation}
\eta_L(t) = L^{-\beta/\nu} \tilde\eta(L^{1/\nu} t)
\label{orderparam_scaled}
\end{equation}
corresponding to a system of box side L. In this expression 
$t=(n-n_c)/n_c$ is the reduced density around the critical point, while
$\nu$ and $\beta$ stand for the critical exponents of the order
parameter and the correlation length, respectively, the later scaling
as $t^{-\nu}$~\cite{Stanley_71}.  In our case we find $\nu$ and
$\beta$ as the optimal values that collapse all curves to a single
$\tilde\eta$ line.  We have found that the best agreement is achieved
for $\nu=0.33$ and $\beta=0.63$.  The upper and lower right panels in
Fig.~(\ref{fig_stripes_054}) show the collapse of the data in the
respective left panels when these values are used. As it can be seen,
the scaling of the data is nicely reproduced, although it is difficult
to accurately determine the exact value of the critical exponents from
the Monte Carlo data.

An interesting result provided by the finite size scaling analysis is
the fact that the values of the critical exponents do not show a
significant dependence on the polarization angle $\alpha$. In this
way, and for $\alpha\geq 0.45$, only the transition density changes
when $\alpha$ varies.  The values of $\nu$ and
$\beta$ derived from our results are compatible with 
the classical 3D Ising universality class, 
$\beta=0.326$ and $\nu=0.630$, associated to the 
$U(1)/Z_2$ symmetry breaking arising when the stripes disappear,
bearing in mind that the critical behavior of a
quantum system in $d$ dimensions is equivalent to that of the
corresponding classical system in $d+1$ dimensions~\cite{Sachdev_11}
The finite size scaling analysis of
the results for different tilting angles allows for the determination
of the gas-stripe transition line, which we summarize in
Table~\ref{Table_gas_Stripe}.  This line turns out to be well fitted
by a curve of the form $n_c r_0^2 = n_0 r_0^2 + a \sin^2 (\alpha -
\alpha_0)$ with $n_0 r_0^2=125.59\pm 3.70, a=18750\pm 2113$ and
$\alpha_0 = 0.6047\pm 0.0052$.

\begin{table}[t]
\begin{tabular}{|c|ccccc|} \hline
$\alpha$ & 0.52 & 0.54 & 0.56 & 0.58 & 0.60 \\  
$n_c r_0^2$ & 260(20) & 205(20) & 160(20) & 140(20) & 125(20) \\  \hline
\end{tabular}
\caption{Tilting angle $\alpha$ and transition
densities $n_c r_0^2$ of the gas-stripe transition.}
\label{Table_gas_Stripe}
\end{table}

It is interesting to notice that the emergence of a stripe phase is
entirely due to the anisotropy of the interaction, but that this
ordering effect is in direct competition with other natural
disordering sources like quantum fluctuations. As a consequence, the
appearance of a stripe phase imposes severe conditions on the system,
in particular on the existence of a threshold density and polarization
angle.  At lower densities than those shown in the table the system
remains in gaseous phase for all values of $\alpha$ up to the collapse
limit.

We end the analysis describing the high density and high polarization
angle region.  We have seen that for low and intermediate values of
$\alpha$ the system remains in solid phase at high densities, while
stripes appear when $\alpha$ is larger than some critical
angle. Consequently, there is a crystal to stripe transition line at
an intermediate region. Getting this transition well characterized is
difficult from the simulation because the system changes from one high
density ordered phase to another.  Still, the different arrangement in
each phase can be observed in the static structure factor $S({\bf
  k})$: the solid phase is characterized by an infinite number of
Bragg peaks located at the characteristic vectors of the reciprocal
lattice, while this is not the case in the stripe phase where ordering
appears in only one direction.

\begin{figure}
\begin{center}
\includegraphics[width=0.49\textwidth]{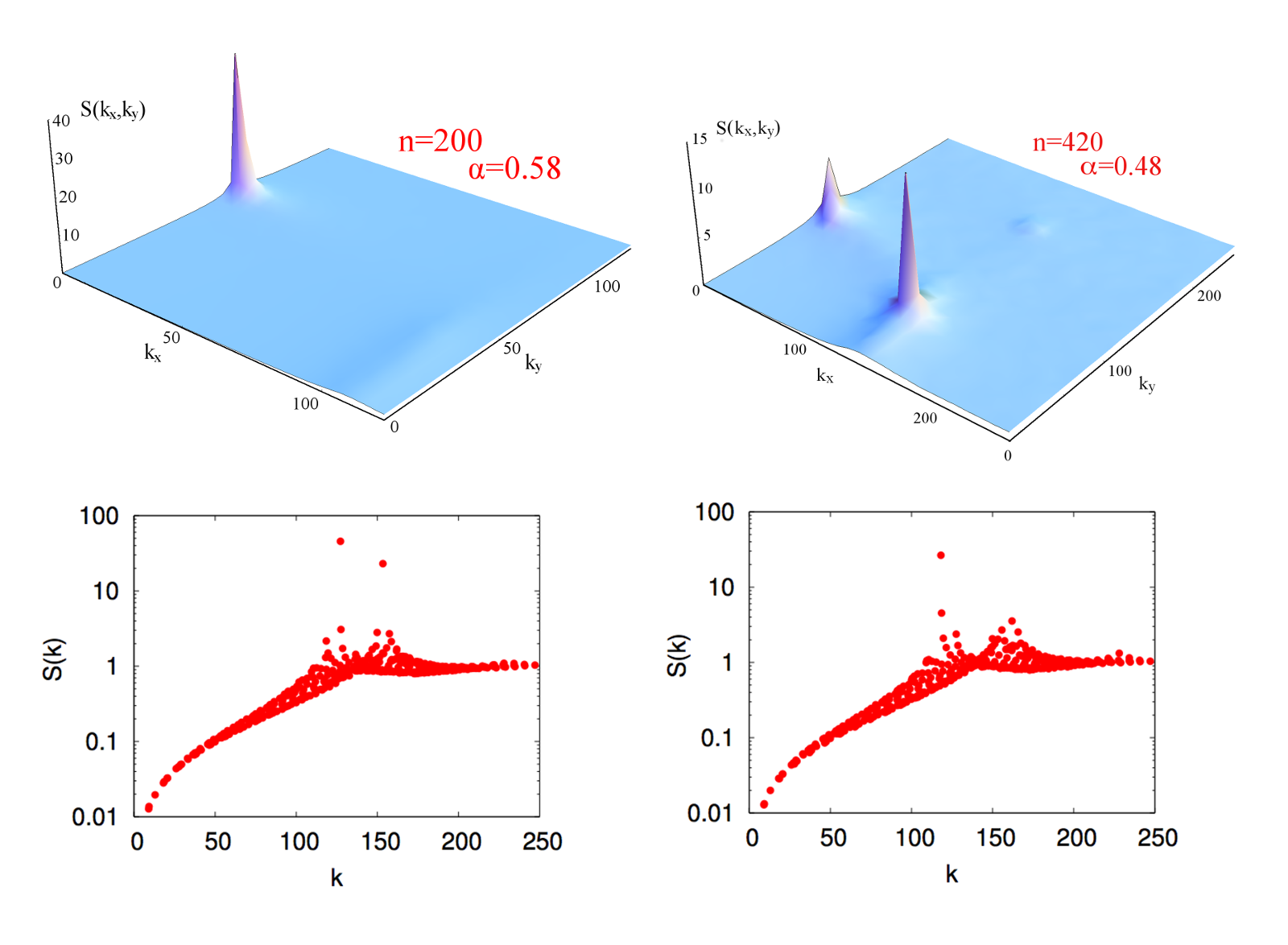}
\caption{(color online). 
Typical form of the static structure factor $S({\bf k})$ of the system in the stripe 
and solid phases (left and right upper panels, respectively).
The lower panels show the projected $S({\bf k})$ for a density $n r_0^2=450$ and
tilting angles $\alpha=0.48$ (left) and $\alpha=0.48125$ (right).
}
\label{fig_Sk_Solid_Stripe}
\end{center}
\end{figure}

The upper panels in Fig.~\ref{fig_Sk_Solid_Stripe} show a typical
example of the full $S({\bf k})$ in the stripe and solid phases as
obtained from our PIGS simulations.
As it can be seen, a second peak is
clearly resolved in the solid phase in comparison with the stripe
phase, where a single Bragg peak associated to the periodicity along
the Y axis is visible. In fact the figure already shows a third peak
in the solid phase that has however much less strength 
but that, together with the other two, put in evidence the structure
of the triangular lattice.  All these peaks are expected to get more
strength when the number of particles in the simulation increases.

In this way, we characterize the stripe to solid phase transition from
the emergence of a second Bragg peak in the simulation, absent in the
former but present in the latter.  In order to describe
the solid-stripe phase
transition we have performed several PIGS simulations with slightly
different values of $\alpha$ and $nr_0^2$, and have
determined through the analysis of the peak structure of $S({\bf k})$
the polarization angle that makes the system lose crystalline order.  The
lower panels in Fig.~\ref{fig_Sk_Solid_Stripe} show the emergence of
the second Bragg peak for a density $n r_0^2=450$ with
$\alpha=0.48$ (left) and $\alpha=0.48125$ (right). The figure shows
the {\em projected} $S({\bf k})$ where the full 2D $S({\bf k})$ is
represented in a single plot with the magnitude of ${\bf k}$ on the X
axis.  
The emergence of the second Bragg peak is quite abrupt and allows
finding the solid to stripe transition point at each fixed density.
The results obtained with this method are shown in
Table~\ref{Table_Solid_Stripe}.  It is remarkable how step the curve
is, thus indicating that the stripe phase seems to be the stable one up
to extremely large densities that are out of reach of our numerical
simulations.

\begin{table}[t]
\begin{tabular}{|c|cccc|} \hline
$n_c r_0^2$ & 450 & 480 & 500 & 550 \\  
$\alpha_c$ & 0.4806(1) & 0.4819(1) & 0.4819(1) & 0.4838(1) \\ \hline
\end{tabular}
\caption{Densities and tilting angles corresponding to 
the solid-stripe transition.}
\label{Table_Solid_Stripe}
\end{table}

An obvious question that arises at this point regards the order of 
the crystal to stripe phase transition. 
In this case we have not detected a smooth decay of the second peak 
of the static structure factor by increasing the polarization
angle of the dipoles. On the contrary, 
the second crystalline Bragg peak suddenly disappears when
a slight change in the tilting angle near the transition point is made,
thus indicating that the crystal to stripe phase transition is probably 
of first order. 

\begin{figure}
\begin{center}
\includegraphics[width=0.49\textwidth]{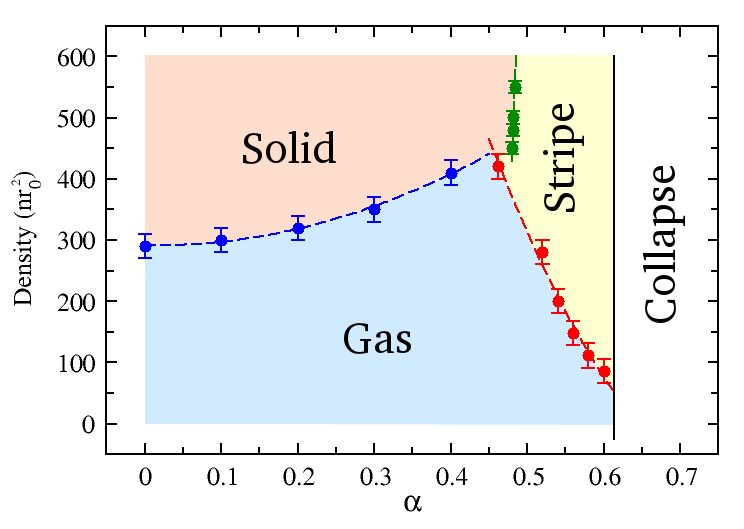}
\caption{(color online). 
Evolution of the order parameter $S_{max}({\bf k})/N$ with the density
for $\alpha=0.1$ (upper left), $\alpha=0.2$ (upper right), $\alpha=0.3$
(lower left) and $\alpha=0.4$ (lower right). }
\label{fig_Phase_Diagram}
\end{center}
\end{figure}

At this point we have covered the whole range of densities and
polarization angles that can be reliably spanned with our simulation
methods. The full phase diagram of the 2D dipolar Bose system is shown
in Fig.~\ref{fig_Phase_Diagram}, where both the simulation points and
the transition lines separating the different phases are depicted and
separated from the collapse region where the system no longer exists.
It is interesting to notice from the phase diagram that there are cuts
at constant density in the range $n r_0^2 \in (290,450)$ where by
increasing the polarization angle from $\alpha=0$ all the way to the
collapse limit one can find the system in solid phase, then in a
gaseous form, to finally jump into the strip phase.  All these changes
are due to a delicate balance between the strength of the interaction
and its anisotropy, which in some sense compete against each other
when the polarization angle increases.

To summarize, using quantum Monte Carlo we have studied the phase
diagram of bosonic tilted dipoles in 2D relying only on the
Hamiltonian of the system.  Our results show that at low densities the
system is in the gas phase. When the density is increased and the
tilting angle is below $\alpha\sim 0.45$ the system undergoes a first
order phase transition and crystallizes. The anisotropy of the
interaction influences the shape of the crystalline lattice by
elongating the fundamental triangle in the direction where the
dipole-dipole interaction is stronger.  Beyond the critical point at
$\alpha\sim 0.45$ a second order phase transition brings the system
from the gas to a stripe phase.  The critical exponents of this second
order transition are essentially independent of the polarization angle
and are compatible with 
the 3D Ising 
universality class within the statistical uncertainty of our
simulations.  Remarkably, our results show that 
for large polarization angles the stripe phase 
can be observed experimentally at densities significantly lower than 
those required to reach the solid phase,
  and without any optical lattice~\cite{ohgoe}.  Finally, at high
densities and large tilting angles the system undergoes a first order
phase transition from the crystal to the stripe phase.  In this
  case, the slope of the transition line is extremely large and
indicates that, due to the anisotropy of the interaction, the crystal
phase of the system can become the most stable one only at extremely
large densities that are nevertheless out of reach of our simulations.

We acknowledge partial financial support from the 
DGI (Spain) Grant No. FIS2011-25275 and  the Generalitat de Catalunya 
Grant No. 2009SGR-1003.

\end{document}